\documentstyle[12pt]{article}

\begin{document}
\begin{flushleft}
Bonn preprint TK-96-17 \\
hep-ph/9608282
\end{flushleft}
\vspace{0.5cm}

\begin{center}
{\huge \bf The role of the pion cloud in the interpretation of
the valence light-cone wavefunction of the nucleon \\}

\vspace{3ex}
\vspace{3ex}
{\large \bf S. D. Bass
\footnote{sbass@pythia.itkp.uni-bonn.de} 
and D. Sch\"utte \footnote{schuette@pythia.itkp.uni-bonn.de} \\}
{\it Institut f\"ur Theoretische Kernphysik, Universit\"at Bonn, \\
Nussallee 14-16, D-53115 Bonn, Germany \\}

\vspace{3ex}
\vspace{3ex}
{\large \bf ABSTRACT \\}
\end{center}
\vspace{3ex}
{
The pion cloud renormalises the light-cone wavefunction of
the nucleon
which is measured in hard, exclusive photon-nucleon reactions.
We discuss the leading twist contributions to high-energy 
exclusive reactions taking into account both the pion cloud
and perturbative QCD physics.
The nucleon's electromagnetic form-factor at high $Q^2$ is 
proportional to the bare nucleon probability $Z$ and the 
cross-sections for hard (real at large angle or deeply virtual)
Compton scattering are proportional to $Z^2$.
Our present knowledge of the pion-nucleon system is consistent
with $Z = 0.7 \pm 0.2$.
If we apply just perturbative QCD to extract a light-cone 
wavefunction directly from these hard exclusive cross-sections, 
then the light-cone wavefunction that we extract measures the 
three valence quarks partially screened 
by the pion cloud of the 
nucleon. 
We discuss how this pion cloud renormalisation effect might 
be understood at the quark level
in terms of the (in-)stability of the perturbative Dirac vacuum 
in low energy QCD.}

\vspace{0.5cm}
\pagebreak
\setlength{\topmargin}{-1.5cm}
\setlength{\textheight}{25.0cm}

\section {Introduction} 

The role of pions in light-cone QCD is a topic of much
theoretical interest [1-7].   
In this paper we explain how the pion cloud renormalises
the valence light-cone wavefunction of the nucleon which 
is measured in the nucleon's electromagnetic form-factor 
at large $Q^2$ and in hard Compton scattering at high energy.
We also discuss how the pion cloud renormalisation of 
high-energy exclusive cross-sections might be understood 
in terms of dynamical symmetry breaking and the 
(in-)stability of the Dirac current-quark vacuum in 
low-energy QCD.

It is well known that the pion cloud, which is associated
with chiral symmetry, plays an important role in the 
phenomenology of nucleon structure [8-11].
Pion cloud effects are present at all momentum scales. 
For example, in nuclear physics the process 
$n \rightarrow p \pi^{-}$ 
offers a simple explanation of the long range part 
of the neutron's electric form-factor \cite{Canjp}. 
The pion cloud of the nucleon renormalises $C=+1$ 
observables like the axial charge of the nucleon 
$g_A^3$ as well as $C=-1$ observables like the nucleon's 
anomalous magnetic moment $\kappa_N$.
In high-energy deep inelastic scattering the process
$p \rightarrow n \pi^+$
generates a non-perturbative component in the nucleon's
sea with an explicit anti-up, anti-down quark asymmetry \cite{AWT83}.
This non-perturbative sea, together with the Pauli 
principle in the nucleon's wavefunction,  explains 
in part \cite{GottTH}
the violation of the Gottfried sum-rule
\cite{Gott} discovered by the NMC \cite{NMC} at CERN --
for a review see Ref. \cite{Shimoda}.
Mesonic effects also play some role in the explanation of
the EMC nuclear effect
\cite{CHLS, ET, PAM}.

In this paper we work on the light-cone and discuss the role 
of the pion cloud in the Fock expansion of the nucleon and 
in high-energy,
exclusive photon-nucleon scattering.
We concentrate on the nucleon's electromagnetic form-factor
and on high-energy Compton scattering.
We discuss how the light-cone wavefunctions which are measured 
in these processes should be interpreted in view of the new 
information we have learnt about the nucleon's internal
structure in unpolarised and polarised (inclusive) deep inelastic 
scattering.

The structure of the paper is as follows.
In Section 2 we start by discussing dynamical chiral symmetry 
breaking (D$ \chi$SB) in low energy QCD.
The emphasis in this Section is on the key physical ideas and 
a supercritical phase transition as a possible explanation of 
the transition from current to constituent quarks.
The picture of low energy QCD which emerges from such a phase
transition is much like the Nambu-Jona-Lasinio model \cite{Weise, NJL}.
In Section 3 we review the theory of how the pion cloud is included 
in the hadronic, light-cone Fock expansion of the nucleon
and how the pion cloud of the nucleon contributes to deep
inelastic scattering.
Our light-cone Fock expansion should provide a unified,
self-consistent approach to both deep inelastic scattering and
high-energy exclusive photon-nucleon reactions -- the subject
of Section 4.
Whilst the pion cloud makes a leading twist contribution to 
deep inelastic structure functions, 
only the bare nucleon (leading hadronic Fock component) makes 
a leading twist contribution to the nucleon's electromagnetic
form-factor at large $Q^2$ and to hard (real at large angle
or deeply virtual) Compton scattering.
At leading twist, the cross sections for high $Q^2$, elastic 
$\gamma p \rightarrow p$ and hard Compton scattering 
are equal to the bare nucleon cross sections multiplied by the
square of 
the bare nucleon probability $Z$, where $Z = 0.7 \pm 0.2$ is 
determined from pion-nucleon physics.
In our present theory of pion cloud effects in deep inelastic
scattering \cite{Shimoda},
the parton model is defined with respect to the bare nucleon 
rather than directly with respect to the physical nucleon.   
The probability to find the physical nucleon in its three-quark
leading Fock state is equal to the bare nucleon probability $Z$
times the probability $P_{3q}$ to find the leading Fock state 
in the bare nucleon.
Section 4 concludes with a discussion how one might best extract
information about the leading Fock state from present and future 
data on 
hard, exclusive processes.
In Section 5 we explain how the pion cloud renormalisation of 
hard exclusive cross sections might be understood
at the quark level
in terms of the vacuum instability picture of D$\chi$SB 
in low-energy QCD
outlined in Section 2.
We discuss how one might construct a light-cone Fock expansion 
that includes both current \cite{Brod91, Brod80} and constituent 
\cite{Wilson, BrodDR} quark degrees of freedom.

\section {Dynamical chiral symmetry breaking in low energy QCD }

QCD is asymptotically free.
At large momentum transfer the running coupling $\alpha_s (Q^2)$
decreases logarithmically with increasing $Q^2$.
The expression for $\alpha_s$ at one loop in perturbation theory is
\begin{equation}
\alpha_s (Q^2) = {4 \pi \over \beta_0 \ln {Q^2 \over \Lambda_{QCD}^2 }},
\end{equation}
where 
$\beta_0 = 11 - {2 \over 3} N_f$ and $N_f$ is the number of flavours.
When $Q^2$ is greater than about 2 GeV$^2$ the running coupling 
$\alpha_s$ is
small enough that one can apply perturbative QCD to calculate the short 
distance (or ``hard")
part of a hadronic scattering process.
The factorisation theorem \cite{Brod89, Coll}
then allows us to write hadronic cross sections
as the convolution
of ``soft" parton distributions 
(in the case of inclusive deep inelastic scattering)
or light-cone wavefunctions
(in the case of high energy exclusive reactions) with a ``hard" scattering
coefficient.
The ``soft" distributions contain all of the information about the structure
of the target -- the long-range bound state dynamics.
They describe a flux of quark and gluon partons into the target independent
``hard" part of the interaction.

Asymptotic freedom also gives us infrared slavery.
The running coupling increases with decreasing resolution $Q^2$. 
Indeed,
the perturbative expression for $\alpha_s$ increases without bound
if we let $Q^2 \rightarrow \Lambda_{QCD}^2$ in Equ.(1).
($\Lambda_{QCD}$ is the infrared Landau scale in QCD.)
On the other hand, perturbation theory is derived assuming that 
the expansion parameter ${\alpha_s \over \pi} \ll 1$.
Physical arguments in non-perturbative QCD suggest that QCD may
undergo a supercritical phase transition at a critical scale $\lambda_c$ 
and that $\alpha_s$ may ``freeze" at the value  $\alpha_s^c = \alpha_s ( \lambda_c)$ 
[26-31].

To understand what happens at this transition 
it is helpful to consider the analogous problem of a
static, large-$Z$, point nucleus in QED \cite{Orsay,Grein1,Grein2}.
There the $1s$ bound state level for the electron falls into the 
negative energy continuum at $Z=137$.
If we attempt to increase $Z$ beyond 137 the point nucleus becomes a
resonance: an electron moves from the Dirac
vacuum to screen the supercritical charge 
which then decays to $Z-1$ with the emission
of a positron.

First, let us consider QCD with just light quarks. If 
the quark itself were to acquire a supercritical charge 
at a critical scale $\lambda_c$, then it would not be 
able to decay into a positive energy bound state together 
with another quark with positive total energy because of 
energy momentum conservation. Instead, the Dirac vacuum 
itself would decay to a new supercritical vacuum state \cite{Orsay}.
Since the vacuum is a scalar, this transition necessarily 
involves the formation of a scalar condensate which 
spontaneously breaks the (near perfect) chiral symmetry 
and yields the massive constituent quark quasi-particles 
of low energy QCD \cite{Miransky, Miran2, Moi}.
We call the phases at scales above and below the critical 
scale $\lambda_c$ the Dirac and Landau phases of QCD respectively.
Perturbative QCD is formulated entirely in the Dirac phase of QCD.
The Dirac vacuum is a highly excited state at scales 
$\mu \leq \lambda_c$ and one must re-quantise the fields
with respect to the new ground state vacuum in the Landau phase of 
the theory.
The Dirac quark of perturbative QCD would freeze out of the theory 
as a dynamical degree of freedom and the running coupling
would freeze at $\alpha_s (\lambda_c)$, which is an infrared, 
unstable fixed point.
The normal ordering mismatch between the zero point energies of 
the scalar vacua in the Dirac and Landau phases of QCD means 
that the quark 
in the low energy
Landau phase feels a uniform, local potential which is manifest 
as the large
mass of the constituent quark quasi-particle.
The chiral dynamics of the Landau phase seem to be well described 
by the Nambu-Jona-Lasinio
model \cite{Weise, NJL, Miransky}.

The freezing of $\alpha_s$ is supported by the analysis of 
infrared induced,
power-behaved contributions 
to hadronic event shapes in $e^+ e^-$ annihilation 
(for example, at
LEP) \cite{Webber}
(see also \cite{Matt}).
Estimates of $\alpha_s^c$ from these experiments range from 
0.6 to 0.8.
(It is also interesting to note that the pre- Sudakov effects
\cite{Li} and pre- chiral symmetry analysis of the high $Q^2$
behaviour of the nucleon's Dirac form factor was consistent with 
freezing of $\alpha_s$
at the (relatively small) value of 0.3 \cite{Lomb}.)

Let us consider this scenario in more detail.
In QCD with both light and heavy quarks there are two types 
of supercritical phase transition associated with a charged 
Dirac vacuum at large coupling $\alpha_s$: 
``static" transitions and ``vacuum" transitions \cite{Orsay,Moi,Three}.
``Static" transitions involve the decay of a heavy quark $q_h$ 
into a light quark $q_l$ together with the formation of a positive 
energy $(q_h {\overline q}_l)$ meson bound state (like the decay 
of the large-$Z$ point nucleus in QED).
These decays may, in part, be responsible for the confinement of 
heavy quarks.
``Vacuum" transitions involve the decay of the fermionic vacuum 
for light-quarks from the Dirac into the Landau phase with the 
formation of a scalar condensate.
Since the $(q_h {\overline q}_l)$ system has a higher reduced mass 
than the $(q_l {\overline q}_l)$ system,
it follows that the static decay of a heavy quark would occur at a 
lower value of $\alpha_s$, or higher value of $Q^2$, than the 
vacuum transition involving light quarks.

To understand the confinement of light-quarks we make the 
simplifying hypothesis that the colour charge at 
$\mu \leq \lambda_c$ is completely screened by the scalar 
condensate. Given that there is a supercritical vacuum 
transition, this hypothesis seems quite reasonable.
If the Dirac quark acquires a supercritical colour charge,
it then becomes a resonance in the negative energy 
continuum
and decays without bound to yield the new vacuum state
containing the scalar condensate and a dynamical colour 
charge that is 
``hidden'' over any ultraviolet cutoff that we may choose 
to regularise the {\it Dirac} Fock space at $\mu \leq \lambda_c$.
What remains is a massive spin ${1 \over 2}$ fermion
which interacts with the condensate.
(The formation of the scalar condensate and consequent mass
generation stabilises the fermion vacuum \cite{Miransky}.)
This massive quasi-particle has a non-dynamical ``passive" 
SU(3)-colour label which is manifest in the SU(3)-colour 
singlet hadronic wavefunctions of the constituent quark model.
In this scenario the colour charge of an isolated quark is 
zero at distances greater than the critical radius
$r_c \sim {1 \over \lambda_c}$ and finite inside $r \leq r_c$.
However, the colour charge is measured by a conserved vector 
current $j^{\mu}$, viz. $ D_{\mu} j^{\mu} = 0$ 
where $D_{\mu}$ is the gauge covariant derivative.
The total colour charge is conserved across the critical 
radius $r_c$.
If nature contained just one isolated quark, the critical 
scale would be infinite ($\lambda_c \rightarrow \infty$) 
so that the colour charge would be completely screened.
In this picture colour confinement is a local phenomenon.
(This effect is essentially the same physics which prevents 
us from having massless charged particles in QED or in the 
Standard Model -- see Refs.\cite{Three, VNG82}.)
For colour singlet hadrons (mesons and baryons) the nett 
colour charge is zero both inside and outside the critical 
radius $r_c$ and the nucleon has finite size.
The colour charge of a given quark is confined to scales 
$\mu \geq \lambda_c$.
This scenario has phenomenological support in the many 
low energy properties of hadrons that can be described 
by the Nambu-Jona-Lasinio model,
which includes chiral symmetry but not dynamical confinement
\cite{Weise, Miran2}.

Clearly, this supercritical confinement scenario differs from 
the confinement 
that is found in pure gluodynamics and in quenched QCD on the 
lattice. 
The importance of chiral symmetry (light quarks) in hadron 
phenomenology poses 
an important question for lattice theorists: 
``Does this gluon induced confinement persist when we relax 
quenching and reduce the light quark mass to its physical value ?'' 
The critical coupling for the Dirac vacuum to become unstable 
increases as we increase the light quark mass (so that fermion vacuum
polarisation is suppressed).
It seems reasonable that the confinement which is observed in pure
gluodynamics may give way to fermion vacuum instability at 
some critical light quark mass.
Whether this critical mass is above or below the physical light 
quark mass is an important question for future lattice calculations.

To conclude this section, we summarise how this physics offers a
possible explanation of the transition from current to constituent 
quarks.
In the Dirac phase of the theory
(at a scale $\mu > \lambda_c$ where $\alpha_s < \alpha_s^c$)
the fermionic degrees of freedom
are the current quarks of perturbative QCD.
In the low-energy Landau phase the fermionic degrees of freedom are 
massive, constituent quark quasi-particles.
The valence quarks in a hadron are the minimal colour-singlet 
combination that enters the hadron's wavefunction:
$q_i {\overline q}_i$ for a meson and $\epsilon_{ijk} q_i q_j q_k$ 
for a baryon.
(Here the subscript $i$ refers to the colour of the quark.)
The sea of quark-antiquark excitations in the Dirac vacuum condense 
in the vacuum transition from the Dirac phase to the Landau phase. 
As a result of this transition the valence current quarks acquire a 
large mass to become the valence constituent quarks of low energy 
QCD. The low energy quark-antiquark condensate is manifest in 
high-energy experiments.
It gives us the infinite number of quark and antiquark partons 
which are observed in the unpolarised, deep inelastic structure 
function at small $x$ and the need for a subtraction in the 
dispersion relation for the total $\gamma p$ cross section \cite{gdh}.
Pions, as the lightest mass excitation of the condensate, should
be included in the nucleon's wavefunction.
As we now discuss, they play an important role in high-energy
photon-nucleon scattering.

\section {The role of the pion cloud in deep inelastic scattering }

Models of the nucleon which include chiral symmetry generally
involve a bare nucleon
and a pion cloud.
The bare nucleon is defined by the SU(3) flavour, SU(2) spin
wavefunction of the three valence constituent quark quasi-particles
(in some confining potential).

Given this picture, the physical nucleon can be viewed on the
light-cone (or in the infinite momentum frame) as the
superposition of the bare nucleon and (in one-meson-approximation)
two-particle
meson-baryon Fock components \cite{Shimoda, WalM, Zoll, SBMa}, viz.
\begin{equation}
|N (p)>_{phys} = Z^{1 \over 2} \Biggl\{ |N (p)>_{bare} 
+ \sum_{M,B}
   \int dy \ dk_T^2 \ g_{0MBN} \ \phi_{MB}(y, k_T) \ 
   | M , B (p, y, k_T) > \Biggr\}.
\end{equation}
Here $Z$ is the bare nucleon probability; $\phi(y, k_T)$ is
the probability amplitude to find the physical nucleon in a
state $|M, B (p, y, k_T)>$  consisting of a meson $M$ and a 
baryon $B$ which carry
light-cone
momentum fractions $y p_+$ and $(1-y) p_+$, and transverse
momentum $k_T$ and $-k_T$ respectively.
Although we work in one-meson-approximation, we include higher
order vertex corrections to the bare coupling $g_{0MBN}$ and
use the dressed hadronic coupling $g_{MBN}^2 = Z g_{0MBN}^2$.
The probability to find the physical nucleon in a state consisting
of the
meson $M$ and baryon $B$
carrying $y$ and $(1-y)$ percent of the physical nucleon's
light-cone momentum $p_+$ is
\begin{equation}
f_{MB} (y) = g_{MBN}^2 \int dk_T^2 | \phi_{MB} (y, k_T ) |^2.
\end{equation}
Conservation of light-cone momentum $p_+$, expressed through the
equation
\begin{equation}
f_{MB} (y) = f_{BM} (1-y),
\end{equation}
provides an important constraint on pion cloud models \cite{Zoll}.
The number of mesons ``in the cloud'' is then
\begin{equation}
<n>_{MB} = <n>_{BM} = \int_0^1 dy f_{MB} (y)
\end{equation}
and the bare nucleon probability is
\begin{equation}
Z = 1 - <n>_{\pi N} - <n>_{\pi \Delta},
\end{equation}
where we now restrict our attention to the pion cloud.

The Fock expansion in Equ.(2) can be used to study the
effect of the pion cloud in deep inelastic scattering
and high-energy exclusive reactions.
At this point we note one important feature of light-cone
perturbation theory:
exchanged quanta are on-mass-shell and off-energy-shell.
In deep inelastic scattering and deeply virtual Compton
scattering
the large momentum squared associated with the exchanged
photon is
$Q^2 = q_T^2$ (strictly speaking $s_{\gamma e^{'}}$) and
$q_{\mu}q^{\mu} = 0$.

Pion cloud contributions to deep inelastic scattering 
are calculated 
via the Sullivan process \cite{Sul72} shown in Fig. 1.
In deep inelastic scattering we measure the inclusive
cross section.
The scattering of the hard photon from the two-particle 
meson-baryon Fock states is a leading twist effect.
The struck hadron is ``shattered'' by the hard photon and,
therefore, does not feel any final state ${\cal O}({1 \over Q^2})$ 
hadronic form-factor.
When we include the Sullivan process, the parton 
distributions of the physical nucleon are obtained
as the convolution of the pionic splitting functions 
$f_{\pi N}(y)$ and $f_{\pi \Delta}(y)$  with the 
parton distributions of the struck hadron,
viz. \cite{Shimoda}:
\begin{eqnarray}
x (q \pm {\overline q})_{N, phys} (x, Q^2) = 
&Z& x (q \pm {\overline q})_{N, bare} (x, Q^2) \\ \nonumber
&+& x \int_x^1 {dy \over y} f_{\pi N} (y) (q \pm {\overline q})_{\pi}
({x \over y}, Q^2) +
x \int_x^1 {dy \over y} f_{N \pi} (y) (q \pm {\overline q})_{N, bare} 
({x \over y}, Q^2) \\ \nonumber
&+&
x \int_x^1 {dy \over y} f_{\pi \Delta}(y) (q \pm {\overline q})_{\pi} 
({x \over y}, Q^2) +
x \int_x^1 {dy \over y} f_{\Delta \pi}(y) 
(q \pm {\overline q})_{\Delta, bare} 
({x \over y}, Q^2)
\end{eqnarray}
and
\begin{eqnarray}
x g_{N, phys} (x, Q^2) =
&Z& x g_{N, bare} (x, Q^2) \\ \nonumber
&+&
x \int_x^1 {dy \over y} f_{\pi N}(y) g_{\pi} ({x \over y}, Q^2) 
+ 
x \int_x^1 {dy \over y} f_{N \pi}(y) g_{N, bare} ({x \over y}, Q^2) 
\\ \nonumber
&+&
x \int_x^1 {dy \over y} f_{\pi \Delta} (y) g_{\pi} ({x \over y}, Q^2) 
+
x \int_x^1 {dy \over y} f_{\Delta \pi} (y) g_{\Delta, bare} 
({x \over y}, Q^2).
\end{eqnarray}
The pionic splitting functions are \cite{Shimoda}:
\begin{equation}
f_{\pi N} (y) =
{3 g_{\pi NN}^2 \over 16 \pi^2 } 
\int_0^{\infty} {dk_T^2 \over (1-y)y}
{ {\cal F}_{\pi N}^2(s_{\pi N}) \over (M_N^2 - s_{\pi N})^2 }
\Biggl( {k_T^2 + y^2 M_N^2 \over 1-y } \Biggr)
\end{equation}
\begin{eqnarray}
f_{\pi \Delta} (y) =
{4 \over 3} {f_{\pi N \Delta}^2 \over 16 m_{\pi}^2 \pi^2 } 
\int_0^{\infty} {dk_T^2 \over (1-y)y} & &
{ {\cal F}_{\pi \Delta}^2(s_{\pi \Delta}) \over (M_N^2 - s_{\pi \Delta})^2 } 
\\ \nonumber
& & { \biggl[ k_T^2 + (M_{\Delta} - (1-y) M_N) \biggr]
  \biggl[ k_T^2 + (M_{\Delta} + (1-y) M_N)^2 \biggr]^2
\over
6 M_{\Delta}^2 (1-y)^3 }.
\end{eqnarray}
Here ${\cal F}_{\pi B}(s_{\pi B})$ is a $\pi BN$ hadronic 
form-factor and
$s_{\pi B}$ is the invariant mass squared of the $\pi B$
intermediate state
\begin{equation}
s_{\pi B} (k_T^2, y) =
{k_T^2 + m_{\pi}^2 \over y} + {k_T^2 + M_B^2 \over 1-y }.
\end{equation}
In phenomenological analyses the form-factor 
${\cal F}_{\pi B}(s_{\pi B})$ is usually written using a
dipole \cite{WalM} or exponential \cite{Zoll} form.
Since $s_{\pi B}$ is invariant under 
($\pi (y) \leftrightarrow B (1-y)$) these pionic splitting 
functions satisfy the $p_+$ conservation equation, Equ.(4).
As a consistency check, note that the physical and bare
nucleon distributions are both correctly normalised to the
number of valence quarks, $N_q$,
in the nucleon:
\begin{equation}
\int_0^1 dx \ (q - {\overline q})_{N, phys} (x, Q^2) = N_q
\end{equation}
and
\begin{equation}
\int_0^1 dx \ (q - {\overline q})_{N, bare} (x, Q^2) = N_q
\end{equation}
where ($N_u =2, N_d =1$) in the proton and ($N_u =1, N_d =2$)
in the neutron.
The process $p \rightarrow n \pi^{+}$ generates an excess 
of anti-down quarks over anti-up quarks in the nucleon's 
wavefunction \cite{AWT83}.
The Sullivan process, together with the Pauli principle 
in the nucleon's wavefunction, offers a simple explanation 
\cite{GottTH}
of the violation of the Gottfried sum-rule observed by the 
NMC \cite{NMC} -- for a review see Ref. \cite{Shimoda}.

There is an important and subtle point to note from Equs.(7,8). 
The quark and gluon distributions of the physical nucleon 
which appear on the left hand side of Equs.(7,8) are the same 
quark and gluon distributions that appear in the operator product 
expansion analysis of deep inelastic scattering.
The parton model distributions are defined via the factorisation 
theorem \cite{Coll} as a flux of quark and gluon partons into the 
hard photon-parton scattering, which is described by the perturbative 
Wilson coefficients. 
{\it These parton model distributions are defined with respect 
to the bare baryon and pion} and not directly with respect to the 
physical nucleon in this approach.
This result will be very important when we discuss high energy
exclusive processes in Section 4.

Anticipating our discussion of exclusive reactions, we will need
the value of the bare nucleon probability $Z$. This quantity is 
determined via Equs.(5,6) and (9,10) by the hardness of the 
pion-nucleon form-factor.
In (equal time) nuclear physics applications the pion nucleon
form-factor is commonly parametrised by a covariant monopole, viz.
\begin{equation}
{\cal F}_{\pi N} (k^2) =
{\Lambda_F^2 - m_{\pi}^2 \over \Lambda_F^2 - k^2}.
\end{equation}
Thomas and Holinde \cite{Hol89} (see also Holinde \cite{Hol94}) 
propose 
that $\Lambda_F = 500-800$MeV (with a preferred value of 730MeV) 
is consistent with pion nucleon phenomenology. This soft pion 
nucleon form-factor is now well accepted in the nuclear physics 
community [47-50].
A recent lattice calculation by Liu and collaborators \cite{Liu95}
gives $\Lambda_F = 750 \pm 140$MeV, which is consistent 
with the Thomas and Holinde result.
Taking $\Lambda_F = 650 \pm 150$MeV in a covariant monopole 
is equivalent to a light-cone form-factor 
${\cal F}_{\pi N}(s_{\pi N})$ 
which corresponds to a bare nucleon probability
\begin{equation}
Z =0.7 \pm 0.2
\end{equation}
where we assume equal hardness of ${\cal F}_{\pi N}$ and 
${\cal F}_{\pi \Delta}$ 
and include an approximate 20\% extra contribution 
from higher mass pseudoscalar and vector mesons \cite{Shimoda}.
In comparison, the value of $Z$ which is calculated in the
Cloudy Bag model
is $Z \simeq 0.5$ for a bag radius $R = 0.8$fm
\cite{ZCBM}.
Some further renormalisation of the exclusive cross-section 
may come from Regge effects which become important when 
$y \rightarrow 1$ in the branching 
process $N \rightarrow N \pi$ and which are not included in
the Fock expansion, Equ.(2) \cite{kopel}.
Given the theoretical uncertainties 
(working in one-meson-approximation, using renormalised couplings)
it may be safer to consider the error on the bare nucleon
probability as a uniform distribution instead of a normal distribution.

\section {The role of the pion cloud in high-energy exclusive reactions}

Following our discussion of pion cloud contributions to deep 
inelastic scattering, we now explain how the pion cloud is 
manifest in high-energy exclusive reactions.
We focus on the nucleon's electromagnetic form-factor $F_N (Q^2)$
at large $Q^2 = q_T^2$ and high-energy Compton scattering.

When we analyse exclusive photon-nucleon reactions at high energy
it is important to take into account both the pion cloud and also
perturbative QCD physics.
We first identify which diagrams contribute to these exclusive
photon-nucleon reactions at leading twist when we work with the 
hadronic Fock expansion in Equ.(2).
Following our discussion of deep inelastic scattering, we then 
apply the perturbative QCD factorisation of Brodsky and Lepage 
\cite{Brod80} to study the hard part of the exclusive reaction.
In this way, we take into account both the physics of the Dirac
and the Landau phases of QCD. (We shall discuss this point further 
in Section 5 below.)

The general rule when discussing pion cloud contributions to 
high-energy photon-nucleon exclusive processes is that diagrams 
which 
involve the flow of large momentum through a hadronic vertex are 
non-leading twist.
This result follows because of the ${\cal O}({1 \over Q^2})$
denominator associated with the propagator of the struck hadron
and also the ${\cal F}_{\pi N}({1 \over Q^2})$ hadronic form-factor 
suppression for the struck baryon (pion) to recombine with the 
spectator pion (baryon) in flight to reconstruct the physical 
nucleon.
The leading twist contribution to the nucleon's electromagnetic
form-factor is
\begin{equation}
F_{N, phys} (Q^2) = Z \ F_{N, bare} (Q^2).
\end{equation}
The bare nucleon probability $Z$ renormalises the large $Q^2$
part of the nucleon's electromagnetic form-factor.
(For an explicit calculation of the higher twist pion cloud 
 contributions to $F_N(Q^2)$ within a particular pion cloud 
 model, see Nikolaev et al.\cite{NNN}.)
Note that $F_{N, phys}(Q^2)$ is the form-factor that one would 
calculate in the Cloudy Bag model, including chiral symmetry,
and 
$F_{N, bare}(Q^2)$ is the form-factor that one would calculate 
retaining only the bare ``MIT core'' (without pions).

The hadronic Fock states in Equ.(2) which contribute to high-energy 
Compton scattering depend on the kinematics.
Let $q_i$ and $q_f$ denote the momenta of the incident and emitted
photons.
Close to the forward direction in ``near-real'' $(q_T \simeq 0$)
Compton scattering, 
where $(q_i - q_f)_{\mu} \ll {\cal O}(\Lambda_F)$, there is an 
explicit pion cloud contribution to the physical cross section 
at leading twist.
This contribution corresponds to diagrams where the incident and
the emitted photons
both couple to the same hadron in Compton scattering from the 
two-particle meson-baryon Fock component
so that no large momentum flows through a hadronic vertex.
In this ``near-real'', ``near-forward'' Compton scattering
one has to consider
explicit 
$\gamma \pi \rightarrow \gamma \pi$, $\gamma N \rightarrow
\gamma N$ and $\gamma \Delta \rightarrow \gamma \Delta$
contributions to the high-energy Compton scattering cross-section.
In Compton scattering away from the forward direction the
two-particle meson-baryon Fock states contribute only at 
higher twist because they involve the flow of large momentum 
through a hadronic vertex.
Such hard processes are real Compton scattering at large angles
and
deeply virtual Compton scattering (dVCS), 
where the nucleon absorbs a large $Q^2 = q_T^2 > {\cal O}(1$GeV$^2)$
photon and radiates a real
photon into the final state.
The cross-section for these hard Compton scattering processes is
\begin{equation}
d \sigma_{N, phys}({\rm hard \ CS}) = Z^2 \ 
d \sigma_{N, bare}({\rm hard \ CS})
\end{equation}
at leading twist.
Having established which hadronic Fock components contribute to 
high-energy elastic photon-nucleon and Compton scattering, 
one can then apply the perturbative QCD analysis of Brodsky 
and Lepage \cite{Brod80}
to $F_N(Q^2)$ and to hard Compton scattering.

The counting rules tell us that the leading twist contribution 
to a hard, exclusive, photon-hadron scattering process is 
given by the hard photon scattering from the valence Fock 
component of 
the hadron involved in the hard scattering \cite{Brod89, Muel, Stoler}.
In their classic work on high-energy exclusive 
reactions \cite{Brod80, Brod89}, Brodsky and Lepage showed that 
perturbative QCD factorisation applies in these 
processes so that the leading twist part of the
nucleon's electromagnetic form-factor can be written
\begin{equation}
F_{N, bare} (Q^2) = 
\int_0^1 [dx]  [dy] \Phi^{*}_{N, bare} (x_i, \mu)
T_H (x_i, y_i, Q, \mu)   \Phi_{N, bare} (y_i,\mu)
\biggl[ 1 + {\cal O}({1 \over Q}) \biggr]
\end{equation}
where $[dx] = dx_1 dx_2 dx_3 \delta(x_1 + x_2 + x_3 - 1)$
and $\mu$ is the factorisation scale.
Here $\Phi_{N, bare} (x_i, \mu)$ is a valence light-cone 
wavefunction which describes the flux of valence quarks 
into the hard scattering, which is described by 
$T_H (x_i, y_i, Q, \mu)$ 
and which may be calculated in perturbative QCD 
\cite{Brod80, Brod89, Li, Muel, Pexc, Hyer}.
This valence wavefunction is defined by the leading term 
in the Fock expansion of the nucleon on the light-cone 
in perturbative QCD --- that is, in the Dirac phase of QCD.

Following Brodsky and Lepage \cite{Brod80}, we introduce a 
partonic
Fock expansion of the bare nucleon in perturbative QCD.
If $p_{\mu}$ and $\lambda$ denote the nucleon's momentum and
helicity respectively, then we write
\begin{equation}
|N (p, \lambda) >_{bare} = 
\sum_{n, \lambda_i} \prod_i^{-}
{dx_i \over \sqrt{x_i}} {d^2 k_{T,i} \over 16 \pi^3}
\psi_{n/N} (x_i, k_{T, i}, \lambda_i)
|n: x_i p_+, {\vec p}_T + {\vec k}_{T, i}, \lambda_i >
\end{equation}
where
\begin{equation}
\prod_i^{-} dx_i d^2 k_{T, i} 
= \prod_i dx_i \delta (\sum_i x_i - 1)
d^2 k_{T, i} 16 \pi^3 \delta (\sum_i k_{T, i}) 
\delta (\sum_i \lambda_i - \lambda).
\end{equation}
In Equ.(19), $\psi_{n/N} (x_i, k_{T, i}, \lambda_i)$ is 
the amplitude for finding the (bare) nucleon in the specific 
Fock state $n$ consisting of partons with momenta
$(x_i p_+, x_i p_T + k_{T, i})$ and helicities $\lambda_i$.
Integrating over the parton's transverse momentum $k_{T, i} < \mu$,
the light-cone wavefunction
\begin{equation}
\Phi (x_i, \mu) =
\int^{\mu} {d^2 k_T \over 16 \pi^3} \psi_{n/N} (x_i, k_{T, i},
\lambda_i)
\end{equation}
describes a flux of partons, collinear up to $k_T < \mu$, into the 
hard scattering 
described by $T_H (x_i, y_i, Q, \mu)$.

In the parton model these light-cone wavefunctions 
are defined with respect to the bare nucleon. 
To see this, consider the (higher-twist) two-particle 
meson-baryon Fock state 
contribution to $F_N (Q^2)$.
The hard scattering takes place either on the baryon with 
the pion in flight or on the pion with the baryon in flight.
The particle in flight does not participate in the hard
scattering process,
whence the factorisation theorem \cite{Brod80, Brod89} 
tells us to use
the light-cone wavefunction of the bare nucleon in the
perturbative QCD part of the analysis of high-energy 
exclusive
scattering.
The probability to find the physical nucleon in its leading
three-quark, valence Fock state is equal to the bare nucleon
probability $Z$ \cite{Cloudy, ZCBM} times the probability 
\begin{equation}
P_{3q}(Q) = \int_0^1 [dx] \Phi^*_{bare}(x_i,Q) \Phi_{bare}(x_i,Q)
\end{equation}
to find the leading Fock state in the bare nucleon 
\cite{Brod80, Brod89}.
The shape of $\Phi_{bare}(x_i,Q)$ determines how the nucleon's
light-cone monentum is distributed among the three valence 
quarks in the leading Fock component.
Given the probability interpretation of these parton model
light-cone wavefunctions, they can also be used to calculate 
the structure function of the bare nucleon
in deep inelastic scattering \cite{Brod89}
\begin{equation}
F_{2,bare} (x,Q^2) =
x \sum_{a = q,g} \int_x^1 {dy  \over y} \sum_{n, \lambda_i}
\int \prod_i^{-} {dz_i  d {\vec k_{T, i}} \over 16 \pi^3}
| \psi_{n/N}^{(Q)} (z_i, {\vec k_{T, i}}, \lambda_i ) |^2
\sum_{b=a} \delta (z_b - y)
C^a({x \over  y}, \alpha_s),
\end{equation}
where we have included the quark charges into the perturbative
coefficients $C^a ({x \over y}, \alpha_s)$.

The light-cone wavefunction of the leading Fock component
in the nucleon 
is calculated from the Fourier transform of the vacuum to 
nucleon matrix element of the nucleon interpolating operator
$ \epsilon^{ijk}
u_{\alpha}^i (z_1) u_{\beta}^j (z_2) d_{\gamma}^k (z_3) $.
(We refer to Chernyak and Zhitnitsky \cite{Chernyak} for technical 
details of this calculation).
The structure of the nucleon enters the calculation
of $\Phi (x_i, Q)$ in the light-cone matrix element
\begin{equation}
<vac | u_{\alpha}^i (z_1) u_{\beta}^j (z_2) 
       d_{\gamma}^k (z_3) | p > \epsilon^{ijk},
\end{equation}
which plays an analogous role to the light-cone correlation 
function in deep inelastic scattering.
Modulo flavour and spin labels, the light-cone wavefunction 
of the leading Fock component can be expanded in terms of the set 
of orthogonal Appell 
polynomials
$A_n (x_i)$ \cite{Brod80, Chernyak}, viz.
\begin{equation}
\Phi (x_i, Q) 
= f_N (Q^2) \ \phi_{as}(x_i) \ \sum_n f_n (Q^2) a_n A_n (x_i).
 \end{equation}
Here $\phi_{as} = 120 x_1 x_2 x_3$ is the asymptotic, 
free-quark-model wavefunction,
$a_n$ are the expansion coefficients,
and $f_N$ and $f_n$ carry the anomalous dimension of the nucleon
interpolating operator
-- that is, they describe the $Q^2$ dependence of the light-cone 
wavefunction.

The normalisation of any theoretical prediction of the light-cone 
wavefunction
$\Phi (x_i, Q)$
depends on the input that one uses for the proton state 
$|N(p,\lambda)>$
and the quark operators in Equ.(24).
(There is no a-priori normalisation of a Bethe-Salpeter amplitude.)
For example, consider the Cloudy Bag model.
The model prediction of the light-cone wavefunction of the leading
Fock component in the physical nucleon is equal to $\sqrt{Z}$ times 
the three-quark, valence wavefunction of the bare (``MIT core'') 
nucleon, which is the light-cone wavefunction used in the parton model. 
(The wavefunction of the physical nucleon measures the
three valence quarks partially screened by the pion cloud.)
For a bag radius $R=0.8$fm \cite{ZCBM} this 
means that 
the pion cloud renormalises the MIT bag model exclusive 
cross-section by a factor of four !
Given a bare nucleon probability $Z=0.7 \pm 0.2$ it is clearly
very important to quantify the extent to which pion corrections
(chiral symmetry)
are included in any given model
calculation of $\Phi (x_i, Q)$ before comparing with data.

Quenched lattice calculations of $\Phi(x_i, Q)$ \cite{Sach}
include some but not all pion loop effects \cite{Tho91, Cohen}.
The situation is somewhat better in QCD sum-rule calculations
\cite{Chernyak, Gari}.
However, it is important to keep in mind that whereas the pion 
cloud renormalises the nucleon mass by about 30\% \cite{Cloudy},
it can renormalise the cross section for hard Compton scattering 
by up to a factor of four.
QCD sum-rule predictions for the unpolarised, hard Compton
scattering cross-section typically differ by a factor of 2-3
\cite{Pexc} 
and, therefore, should not be distinguished by comparison with 
the absolute, measured cross-section alone.

The shape of the spin-independent, parton-model wavefunction
for the three valence quarks without pionic dressing
($\phi_{as}(x_i) \sum_n f_n(Q^2) a_n A_n (x_i)$ in Equ.(25))
is best determined from experiment by a maximum likelihood 
fit to $Z$ independent ratios of unpolarised, hard, exclusive 
observables such as:
\begin{eqnarray}
R_1 &=& 
{ d \sigma ({\rm dVCS}, \theta) \over 
  d \sigma ({\rm dVCS}, \theta=90^{0}) } \\
R_2 &=&
{ d \sigma ({\rm dVCS}, \theta) \over F_N^2 (Q^2) }.
\end{eqnarray}
In performing these fits, it is important to make sure that
the observables in the numerator and denominator are described
self-consistently using the same factorisation scheme and 
also at the same factorisation scale (to eliminate factors of $f_N$).
The sensitivity of exclusive cross-sections to the normalisation of 
light-cone wavefunctions
has been stressed previously by Hyer \cite{Hyer}.
Given the fairly large uncertainty in the bare nucleon probability,
one should compare the predictions of lattice \cite{Sach}, 
QCD sum-rule \cite{Chernyak, Gari} and diquark \cite{Kroll} models
for the shape of light-cone wavefunctions
with $Z$ independent ratios of hard, exclusive cross-sections 
instead of the absolute cross-sections.

There has been much theoretical work in recent years aimed at
understanding the EMC spin effect \cite{Spin} in polarised 
deep inelastic scattering.
Two important theoretical results are the role of the axial
anomaly in spin dependent processes \cite{etar},
and possible contributions from QCD background fields \cite{Spin1,Spin2}.
The anomaly can induce a contact interaction between a hard photon
and the background field (on non-perturbative vacuum) \cite{Spin1,jm},
which has no Fock representation in perturbation theory.
This interaction is, in general, leading twist and has the potential
to screen the spin of the quarks at large $x$ in the spin dependent
structure function $g_1$ \cite{Spin1}.
The phenomenology of this effect is that any $C=+1$ spin observable
is, in principle, subject to a significant violation of Zweig's rule.
This Zweig's rule violation has the potential to modify the shape as
well as the normalisation of 
the spin dependent light-cone wavfunction which is measured in hard,
exclusive scattering.
How might one try to isolate such a Zweig's rule violation ?
Experimentally, this requires a comparison between the light-cone
wavefunction which is measured in $C=+1$ and $C=-1$ spin observables.
(Such $C=-1$ spin observables are found in parity-violating, hard 
 Compton scattering.)
To isolate the possible Zweig's rule violation in theoretical
calculations, 
one needs to introduce a switch in the QCD sum-rule calculation with 
which to turn on and off the effect of the anomaly  --- possibly
along the lines suggested by
Narison, Shore and Venziano in polarised deep inelastic scattering
\cite{venez}.

\section {D$\chi$SB, confinement and light-cone QCD }

In Sections 3 and 4 we introduced a two-stage Fock expansion
to describe high-energy exclusive reactions  in light-cone 
QCD. At the first stage we introduced pions explicitly and 
at the second stage we introduced (perturbative) quark and 
gluon degrees of freedom. It is worthwhile to stop and ask 
whether one could expand the physical nucleon directly in terms
of perturbative quark and gluon Fock components and 
not lose any of the physics.

In light-cone perturbation theory all particles are on-mass-shell 
and potentially off-energy-shell.
For a given particle
\begin{equation}
k_+ = {m^2 + k_T^2 \over k_-} \geq 0,
\end{equation}
where the equality holds only for massless particles. 
When we quantise perturbative QCD on the  light-cone the current
quark has a well defined mass. 
We can set $k^2 = m^2$ so that $k_+ > 0$ and successfully apply 
light-cone perturbation 
theory \cite{Brod91}.
Confinement becomes important at large coupling --- at which point
the Dirac quark has no well defined mass shell in (equal-time) QCD.
If the Dirac quark becomes a supercritical resonance,  
then $k_+$ develops an imaginary part and the $k_+ \geq 0$ 
constraint no longer applies. 
(Even the real part of $k_+$ can go negative.)
Conventional light-cone perturbation theory \cite{Brod91} does 
not apply in an unstable vacuum.
Motivated by our discussion in Section 2, this problem is cured
if we work with massive constituent quarks with non-dynamical 
colour, such as in the Nambu-Jona-Lasinio model \cite{Weise, NJL}.
Of course, we need to remember that colour is confined.
This means that the bare nucleon (composed of constituent
quarks) is the intermediate step between the physical 
nucleon and perturbative QCD in the Fock expansion of
the nucleon.
We do not introduce a perturbative QCD Fock expansion of 
the constituent 
quark as the intermediate step because the constituent quark 
has no meaning
when considered in isolation. (There are no free quarks.)

Finally, we outline how the chiral symmetry renormalisation of 
exclusive cross-sections might be understood at the quark level
in terms of an unstable Dirac vacuum.
Fradkin and collaborators \cite{Frad1} have derived the transition 
probabilities 
for exclusive reactions in QED with an unstable vacuum, which they 
include via an external field. 
Here, one needs to consider contributions to exclusive cross-sections 
where the vacuum ``decays'' during the exclusive scattering 
with the emission of $e^- e^+$ pairs.
(Remember that the $k_+ \geq 0$ constraint no longer applies once we
turn on vacuum instability.)
It follows that
\begin{equation}
p_V = |< vac (out) \ | \ vac (in) >|^2 < 1
\end{equation}
in an unstable vacuum.
The creation (annihilation) operator for out-state electrons in QED
with an unstable vacuum is
a non-trivial, linear
 superposition of the creation (annihilation) operator
for in-state electrons and the annihilation (creation) operator for
in-state positrons \cite{Frad1}.
Vacuum instability mixes the Fock components of the in- and out- 
state 
wavefunctions.

Turning now to QCD, this effect is clearly not important at short 
distances where $\alpha_s$ is small -- the ``hard" part of the 
exclusive reaction. 
It is potentially very important for our interpretation of the 
``soft" light-cone wavefunctions if the Dirac quark is allowed 
to undergo a supercritical decay between light-cone times
$\tau \rightarrow - \infty$ and $\tau \rightarrow + \infty$ in
the exclusive reaction.
In this case, what we call a quark in the in-state is modified
by the charged vacuum en route to what we call a quark in the 
out-state of the scattering process.
The exclusive cross-section is then suppressed with respect to 
the cross section that we would predict if we assume that the 
Dirac vacuum is stable at all scales.
This suppression is identified with the bare nucleon probability 
in the vacuum instability mechanism for dynamical symmetry breaking 
outlined in  Section 2
(which suggests a possible approach how one might calculate $Z$
in an eventual solution to non-perturbative QCD).
The supercritical decay is ``static'' for a heavy quark and ``vacuum''
for a light quark.
The flavour-SU(N) and spin-SU(2) valence wavefunction of a given 
hadron determines the relative importance of ``static'' and ``vacuum''
transitions at large coupling
$\alpha_s$
and, therefore, the hadron dependence of the bare hadron probability $Z$.

\section {Conclusions }

Recent experiments in inclusive, deep inelastic scattering 
have shed new information on the role of the pion cloud 
(and possible background field effects)
in the structure of the nucleon.
These discoveries are also important to our interpretation
of the light-cone wavefunctions which are measured in hard, 
exclusive scattering.
The pion cloud renormalises the cross-sections for high-energy 
exclusive photon-nucleon processes. 
The nucleon's electromagnetic form-factor $F_N (Q^2)$ at large 
$Q^2$
is renormalised by the bare nucleon probability $Z$ and the cross 
section for hard Compton scattering is renormalised by $Z^2$.
The nucleon's valence light-cone wavefunction in the parton model
is defined with respect to the bare nucleon  (the MIT core in the 
Cloudy Bag model of the nucleon). 
The probability to find the physical nucleon in its leading 
three-quark, valence Fock state is equal to the bare nucleon 
probability $Z$ \cite{Cloudy, ZCBM} times the probability $P_{3q}$ 
to find the leading Fock state in the bare nucleon \cite{Brod80, Brod89}.
Given the fairly large uncertainty on the value of $Z$ (nearly a
factor of two), it is important to compare the predictions of
various models (lattice, QCD sum-rules, diquark) with $Z$ independent
ratios of hard, exclusive observables
rather than the absolute cross-sections.
If we apply just perturbative QCD to extract a light-cone 
wavefunction directly from the cross-sections for hard real and
deeply virtual Compton scattering data, 
then the light-cone wavefunction that we extract has 
the interpretation that it measures the three valence quarks 
partially screened by the pion cloud of the nucleon. 
Dirac vacuum instability in low-energy QCD offers a possible 
quark level explanation of the pion cloud renormalisation of 
hard, exclusive photon-nucleon
cross-sections.

\vspace{2.0cm}

{\large \bf Acknowledgements \\ }

It is a pleasure to thank S. J. Brodsky, A. C. Kalloniatis, 
B. Kopeliovich, O. Nachtmann, H. Petry, A. W. Thomas and 
W. Weise for helpful discussions on various aspects of 
this work. 
We gratefully acknowledge the support of a 
Research Fellowship of the Alexander von Humboldt Foundation 
(S.D.B.) and the Deutsche Forschungsgemeinschaft (D.S.).



\pagebreak

\input FEYNMAN
\begin{center}
\begin{picture}(30000,30000)

\drawline\photon[\NW\CURLY](5000,15000)[8]
\put(-3000, 18000){$\gamma^{*}(q)$}
\drawline\fermion[\E\REG](\photonfrontx,\photonfronty)[5000]
\drawarrow[\E\ATTIP](\pbackx, \pbacky)
\drawline\fermion[\N\REG](\photonfrontx, \photonfronty)[750]
\drawline\fermion[\E\REG](\fermionbackx, \fermionbacky)[5000]
\drawarrow[\E\ATTIP](\pbackx,\pbacky)
\drawline\fermion[\S\REG](\photonfrontx, \photonfronty)[750]
\drawline\fermion[\E\REG](\fermionbackx, \fermionbacky)[5000]
\drawarrow[\E\ATTIP](\pbackx, \pbacky)
\put(\photonfrontx,\photonfronty){\circle*{1500}}
\drawline\scalar[\SW\REG](\photonfrontx, \photonfronty)[4]
\put(\pmidx, \pmidy){\,\,\,\,{\bf M}$(y, k_T)$}
\thicklines\drawline\fermion[\W\REG](\scalarbackx, \scalarbacky)[5000]
\put(-5000, 7000){$N(p)$}
\drawarrow[\E\ATTIP](\pmidx, \pmidy)
\thicklines\drawline\fermion[\SE\REG](\scalarbackx, \scalarbacky)[5000]
\drawarrow[\SE\ATTIP](\pbackx, \pbacky)
\put(\pmidx, \pmidy){\,\,{\bf B}$(1-y, -k_T)$}
\thinlines

\drawline\photon[\NW\CURLY](25000,15000)[8]
\put(17000, 18000){$\gamma^{*}(q)$}
\drawline\fermion[\E\REG](\photonfrontx,\photonfronty)[5000]
\drawarrow[\E\ATTIP](\pbackx, \pbacky)
\drawline\fermion[\N\REG](\photonfrontx, \photonfronty)[750]
\drawline\fermion[\E\REG](\fermionbackx, \fermionbacky)[5000]
\drawarrow[\E\ATTIP](\pbackx,\pbacky)
\drawline\fermion[\S\REG](\photonfrontx, \photonfronty)[750]
\drawline\fermion[\E\REG](\fermionbackx, \fermionbacky)[5000]
\drawarrow[\E\ATTIP](\pbackx, \pbacky)
\put(\photonfrontx,\photonfronty){\circle*{1500}}
\thicklines\drawline\fermion[\SW\REG](\photonfrontx, \photonfronty)[8000]
\put(\pmidx, \pmidy){\,\,\,\,{\bf B}$(y, k_T)$}
\thinlines\drawline\scalar[\SE\REG](\fermionbackx, \fermionbacky)[3]
\drawarrow[\SE\ATTIP](\pbackx, \pbacky)
\put(\pmidx, \pmidy){\,\,{\bf M}$(1-y, -k_T)$}
\thicklines\drawline\fermion[\W\REG](\scalarfrontx, \scalarfronty)[5000]
\drawarrow[\E\ATTIP](\pmidx, \pmidy)
\put(15000,7000){$N(p)$}

\end{picture}
\end{center}

\begin{center}
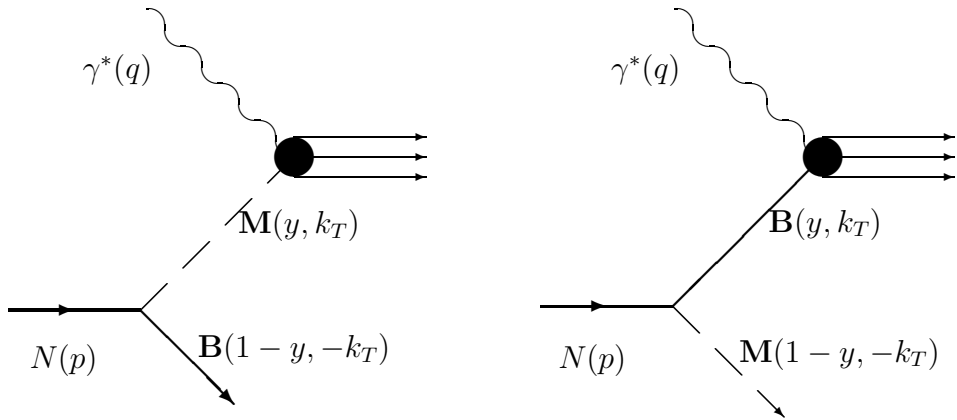
Fig.1: Two-particle meson-baryon contributions to deep inelastic
scattering
\end{center}


\begin{thebibliography}{99}

\bibitem{Wilson}
K. G. Wilson  et al., Phys. Rev. D49 (1994) 6720.
%
\bibitem{Glazek}
Various contributions in {\it Theory of hadrons and light-cone QCD},
ed. St.D. Glazek (World Scientific, 1995). 
%
\bibitem{Wern}
T. Heinzl, S. Krushe and E. Werner, Z. Physik C56 (1992) 415.
%
\bibitem{DRob}
D. G. Robertson, Phys. Rev. D47 (1993) 2549.
%
\bibitem{Yama}
Y. Kim, S. Tsujimaru and K. Yamawaki, Phys. Rev. Letts. 74 (1995) 4771.
%
\bibitem{Burk}
M. Burkardt, hep-ph 9505259, to appear in Adv. Nucl. Phys.
%
\bibitem{BrodDR}
S. J. Brodsky and D. G. Robertson, in {\it Confinement Physics},
eds. S. D. Bass and P. A. M. Guichon (Editions Frontieres, 1996).
%
\bibitem{Ericson}
T. E. O. Ericson and W. Weise, {\it Pions and Nuclei}, Oxford UP (1988).
%
\bibitem{Cloudy}
A. W. Thomas, Adv. Nucl. Phys. 13 (1984) 1.
%
\bibitem{pion}
A. W. Thomas, Nucl. Phys. A518 (1990) 186.
%
\bibitem{Weise} U. Vogl and W. Weise, Prog. Part. Nucl. Phys. 26 (1991) 195.
%
\bibitem{Canjp}
S. Th\'eberge, G. A. Miller and A. W. Thomas, 
Can. J. Phys. 60 (1982) 59.
%
\bibitem{AWT83}
A. W. Thomas, Phys. Lett. B126 (1983) 97.
%
\bibitem{GottTH}
E. M. Henley and G. A. Miller, Phys. Lett. B251 (1990) 453; \\
S. Kumano and J. T. Londergan, Phys. Rev. D44 (1991) 717; \\
A. Signal, A. W. Schreiber and A. W. Thomas,
Mod. Phys. Lett. A6 (1991) 271.
%
\bibitem{Gott}
K. Gottfried, Phys. Rev. Lett. 18 (1967) 1174.
%
\bibitem{NMC}
The New Muon  Collaboration, P. Amaudruz et al., 
Phys. Rev. Lett. 66 (1991) 2712.
%
\bibitem{Shimoda}
A. W. Thomas and W. Melnitchouk, in {\it New  Frontiers in Nuclear
Physics},
eds. S. Homma, Y. Akaishi and M. Wada (World Scientific, Singapore, 1993),
pp. 41-106.
%
\bibitem{CHLS}
C. H. Llewellyn Smith, Phys. Lett. B128 (1983) 107.
%
\bibitem{ET}
M. Ericson and A. W. Thomas, Phys. Lett. B128 (1983) 112.
%
\bibitem{PAM}
A. W. Thomas, A. Michels, A. W. Schreiber and P. A. M. Guichon,
Phys. Lett. B233 (1989) 43; \\
K. Saito and A. W. Thomas, Nucl. Phys. A574 (1994) 639.
%
\bibitem{NJL} 
Y. Nambu and G. Jona-Lasinio, Phys. Rev. 122 (1961) 345.
%
\bibitem{Brod91}
S. J. Brodsky and H-C. Pauli, Lecture Notes
in Physics, Vol. 396 (Springer-Verlag, 1991).
%
\bibitem{Brod80}
G. P. Lepage and S. J. Brodsky, Phys. Rev. D22 (1980) 2157.
%
\bibitem{Brod89}
S. J. Brodsky and G. P. Lepage, in {\it Perturbative Quantum
Chromodynamics}, ed. A. Mueller, World Scientific (1989).
%
\bibitem{Coll}
J. C. Collins, D. E. Soper and G. Sterman, in {\it Perturbative
Quantum Chromodynamics}, ed. A. Mueller (World Scientific, 1989); \\
G. Sterman, hep-ph 9606312 (1996).
%
\bibitem{Miransky}
V. A. Miransky, {\it Dynamical symmetry breaking in quantum field
theories},
World Scientific (1993).
%
\bibitem{Miran2}
P. I. Fomin, V. P. Gusynin, V. A. Miransky and Yu. A. Sitenko, 
``Dynamical symmetry breaking and particle mass generation in gauge field 
theories",
Riv. Nuovo Cimento 6 (1983) 1.
%
\bibitem{Man}
J. E. Mandula, Phys. Lett. B67 (1977) 175; \\
A. J. G. Hey, D. Horn and J. E. Mandula, Phys. Lett. B80 (1978) 90. 
%
\bibitem{Lund}
V. N. Gribov, Physica Scripta T15 (1987) 164; \\
Lund preprint LU TP 91/7 (May 1991), unpublished. 
%
\bibitem{Orsay}
V. N. Gribov, 
Orsay lectures LPTHE Orsay 92/60 (June 1993) and LPTHE Orsay 94/20
(Feb. 1994).
%
\bibitem{Moi}
S. D. Bass, in {\it Confinement Physics}, 
eds. S. D. Bass and P. A. M. Guichon (Editions Frontieres, 1996).
%
\bibitem{Grein1} 
I. Pomeranchuk and Ya. Smorodinsky, J. Fiz. USSR 9 (1945) 97; \\
W. Pieper and W. Greiner, Z Physik 218 (1969) 327; \\
Ya. B. Zeldovich and V. S. Popov, Uspekhi Fiz. Nauk. 105 (1971) 4.
%
\bibitem{Grein2}
W. Greiner, B. M{\"u}ller and J. Rafelski, 
{\it Quantum electrodynamics of
strong fields}, Springer-Verlag (1985)
%
\bibitem{Webber}
Yu. L. Dokshitzer and B. R. Webber, Phys. Lett. B352 (1995) 451 \\
Yu. L. Dokshitzer, G. Marchesini and B. R. Webber, hep-ph/9512336 (1995).
%
\bibitem{Matt}
A. C. Mattingly and P. M. Stevenson, Phys. Rev. D49 (1994) 437
%
\bibitem{Li}
H-N. Li and G. Sterman, Nucl. Phys. B 381 (1992) 129;
H-N. Li, Phys. Rev. D48 (1993) 4243.
%
\bibitem{Lomb}
C-R. Ji, A. F. Sill and R. M. Lombard-Nelson, Phys. Rev. D36 (1987) 165
%
\bibitem{Three}
S. D. Bass and A. W. Thomas, Mod. Phys. Lett. A11 (1996) 339
%
\bibitem{VNG82}
V. N. Gribov, Nucl. Phys. B206 (1982) 103.
%
\bibitem{gdh} 
S. D. Bass, Z Physik A355 (1996) 77.
%
\bibitem{WalM}
W. Melnitchouk and A. W. Thomas, Phys. Rev. D47 (1993) 3794.
%
\bibitem{Zoll}
V. R. Zoller, Z Physik C54 (1992) 425.
%
\bibitem{SBMa}
S. J. Brodsky and B-Q. Ma, Phys. Lett. B381 (1996) 317.
%
\bibitem{Sul72}
J. D. Sullivan, Phys. Rev. D5 (1972) 1732.
%
\bibitem{Hol89}
A. W. Thomas and K. Holinde, Phys. Rev. Lett. 63 (1989) 2025.
%
\bibitem{Hol94}
K. Holinde, in Proc. {\it Physics with GeV particle beams},
eds. H. Machner and K. Sistemich (World Scientific (1995), pp.285-296.
%
\bibitem{Schu88}
D. Sch\"utte and A. Tillemans, Phys. Lett. B206 (1988) 1.
%
\bibitem{Hol90}
K. Holinde and A. W. Thomas, Phys. Rev. C42 (1990) R1195; \\
G. Jansen, K. Holinde and J. Speth, Phys. Rev. Lett. 73 (1994) 1332.
%
\bibitem{Brown}
G. E. Brown, M. Buballa, Zi Bang Li and J. Wambach, 
Nucl. Phys. A593 (1995) 295.
%
\bibitem{Fra96}
W. Koepf, L. L. Frankfurt and M. Strikman, 
Phys. Rev. D53 (1996) 2586.
%
\bibitem{Liu95}
K. F. Liu, S. J. Dong, T. Draper and W. Wilcox, 
Phys. Rev. Lett. 74 (1995) 2172.
%
\bibitem{ZCBM}
A. W. Thomas, S. Th\'eberge and G. A. Miller, Phys. Rev. D24 (1981) 216.
%
\bibitem{kopel}
B.Z. Kopeliovich, B. Povh, I.K. Potashnikova, hep-ph/9601291,
Z Physik C, in press
%
\bibitem{NNN}
N. N. Nikolaev, A. Szczurek, J. Speth and V. R. Zoller, 
Z Physik A349 (1994) 59.
%
\bibitem{Muel}
A. H. Mueller, in {\it The Heart of the Matter}, 
Proc. VI Rencontres de Blois, 
eds. J-F. Mathiot and J. Tran Thanh Van,
Editions Frontieres (1994).
%
\bibitem{Stoler}
P. Stoler, Phys. Reports 226 (1993) 103.
%
\bibitem{Pexc}
G. R. Farrar and H. Zhang, Phys. Rev. D41 (1990) 3348; \\
A. S. Kronfield and B. Nizi\'c, Phys. Rev. D44 (1991) 3445.
%
\bibitem{Hyer}
T. Hyer, Phys. Rev. D47 (1993) 3875; \\
T. Hyer, Ph.D. thesis, SLAC-report-441.
%
\bibitem{Chernyak}
V. L. Chernyak and I. R. Zhitnitsky, Phys. Rept. 112 (1984) 173; 
Nucl. Phys. B246 (1984) 52; Z Physik C42 (1989) 569; \\
I. R. Zhitnitsky, A. A. Oglobin and V. L. Chernyak, Yad. Fiz. 48
(1988) 841.
%
\bibitem{Kroll}
P. Kroll, M. Schurmann and W. Schweiger, Int. J. Mod. Phys. A6 (1991)
4107; \\
P. Kroll, M. Schurmann and P.A.M. Guichon, Nucl. Phys. A598 (1996)
435.
%
\bibitem{Sach}
G. Martinelli and C. T. Sachrajda, Phys. Lett. B217 (1989) 319; \\
A. S. Kronfeld and D. M. Photiadis, Phys. Rev. D31 (1985) 2939.
%
\bibitem{Tho91}
A. W. Thomas, Aust. J. Phys. 44 (1991) 173.
%
\bibitem{Cohen}
T. D. Cohen and  D. B. Leinweber, 
Comm. Nucl. Part. Phys. 21 (1993) 137.
%
\bibitem{Gari}
M. Gari and N. G. Stefanis, Phys. Lett. B175 (1986) 462;  \\
I. D. King and C. T. Sachrajda, Nucl. Phys. B279 (1987) 785.
%
\bibitem{Spin}
The European Muon Collaboration, J. Ashman et al., Phys. Lett. B206
(1988) 364; Nucl. Phys. B328 (1990) 1.
%
\bibitem{etar}
A.V. Efremov and O.V. Teryaev, Dubna preprint E2-88-287 (1988); \\
G. Altarelli and G.G. Ross, Phys. Lett. B212 (1988) 391; \\
R.D. Carlitz, J.C. Collins and A.H. Mueller, Phys. Lett. B214 (1988)
229.
%
\bibitem{Spin1}
S. D. Bass, Phys. Lett. B367 (1996) 335.
%
\bibitem{Spin2}
O. Nachtmann, in {\it Confinement Physics}, eds. S.D. Bass and
P.A.M. Guichon (Editions Frontieres, 1996).
%
\bibitem{jm}
R.L. Jaffe and A. Manohar, Nucl. Phys. B337 (1990) 509.
%
\bibitem{venez}
G. Veneziano, Mod. Phys. Lett. A4 (1989) 1605; \\
G.M. Shore and G. Veneziano, Nucl. Phys. B381 (1992) 23; \\
S. Narison, G. M. Shore and G. Venziano, Nucl. Phys. B433 (1995) 209.
%
\bibitem{Frad1}
E.S. Fradkin, D.M. Gitman and S.M. Shvartsman,
{\it Quantum electrodynamics with unstable vacuum}, Springer-Verlag (1991).
%
\end{thebibliography}
\end{document}